\documentclass[
superscriptaddress,
 amsmath,amssymb,
 aps,
reprint
]{revtex4-1}

\usepackage{color}
\usepackage{graphicx}
\usepackage{dcolumn}
\usepackage{bm}
\usepackage{epstopdf}
\usepackage{gensymb}

\usepackage{float}

\begin{document}


\title{Effect of Pt vacancies on magnetotransport of Weyl semimetal candidate GdPtSb epitaxial films}

\author{Dongxue Du}
\affiliation{Materials Science and Engineering, University of Wisconsin-Madison, Madison, WI 53706}

\author{Laxman Raju Thoutam}
\affiliation{Amrita School of Nanosciences and Molecular Medicine, Amrita Vishwa Vidyapeetham, Ponekkara, Kochi 682041, India}
\affiliation{Department of Chemical Engineering and Materials Science, University of Minnesota -- Twin Cities}

\author{Konrad T. Genser}
\affiliation{Department of Physics and Astronomy, Rutgers University}

\author{Chenyu Zhang}
\affiliation{Materials Science and Engineering, University of Wisconsin-Madison, Madison, WI 53706}
\author{Karin M. Rabe}
\affiliation{Department of Physics and Astronomy, Rutgers University}

\author{Bharat Jalan}
\affiliation{Department of Chemical Engineering and Materials Science, University of Minnesota -- Twin Cities}
\author{Paul M. Voyles}
\affiliation{Materials Science and Engineering, University of Wisconsin-Madison, Madison, WI 53706}
\author{Jason K. Kawasaki}
\affiliation{Materials Science and Engineering, University of Wisconsin-Madison, Madison, WI 53706}
\email{jkawasaki@wisc.edu}

\date{\today}
\begin{abstract}

We examine the effects of Pt vacancies on the magnetotransport properties of Weyl semimetal candidate GdPtSb films, grown by molecular beam epitaxy on c-plane sapphire. Rutherford backscattering spectrometry (RBS) and x-ray diffraction measurements suggest that phase pure GdPt$_{x}$Sb films can accommodate up to $15\%$ Pt vacancies ($x=0.85$), which act as acceptors as measured by Hall effect. Two classes of electrical transport behavior are observed. Pt-deficient films display a metallic temperature dependent resistivity (d$\rho$/dT$>$0). The longitudinal magnetoresistance (LMR, magnetic field $\mathbf{B}$ parallel to electric field $\mathbf{E}$) is more negative than transverse magnetoresistance (TMR, $\mathbf{B} \perp \mathbf{E}$), consistent with the expected chiral anomaly for a Weyl semimetal. The combination of Pt-vacancy disorder and doping away from the expected Weyl nodes; however, suggests conductivity fluctuations may explain the negative LMR rather than chiral anomaly. Samples closer to stoichiometry display the opposite behavior: semiconductor-like resistivity (d$\rho$/dT$<$0) and more negative transverse magnetoresistance than longitudinal magnetoresistance. Hysteresis and other nonlinearities in the low field Hall effect and magnetoresistance suggest that spin disorder scattering, and possible topological Hall effect, may dominate the near stoichiometric samples. Our findings highlight the complications of transport-based identification of Weyl nodes, but point to possible topological spin textures in GdPtSb.

\end{abstract}

\maketitle

\section{Introduction}

The lanthanide half Heusler compounds $Ln$PtBi and $Ln$PtSb are attractive due to their tunable topological and magnetic properties as functions of lanthanide substitution \cite{nakajima2015topological, lin2010half,gofryk2011magnetic}, strain \cite{lin2010half}, and strain gradients \cite{du2021epitaxy}. Compounds in this family of materials were among the first identified as zero bandgap topological semimetals via density functional theory (DFT) \cite{lin2010half} with confirmation for LuPtSb \cite{logan2016observation}, LuPtBi \cite{liu2016observation} and YPtBi \cite{liu2016observation} by angle-resolved photoemission spectroscopy (ARPES) measurements \cite{liu2016observation,logan2016observation}.

More recently, bandstructure calculations and magnetotransport measurements suggest that GdPtBi \cite{hirschberger2016chiral, shekhar2018anomalous}, TbPtBi \cite{chen2020chiral}, HoPtBi \cite{chen2020chiral}, and ErPtBi \cite{chen2020chiral} compounds are magnetic-field-induced Weyl semimetals. In these materials, magnetic field, either directly through Zeeman splitting or indirectly through exchange splitting from field-induced magnetization, is expected to lift the degeneracy of quadratic bands that touch at $\Gamma$, to create pairs of Weyl nodes \cite{hirschberger2016chiral, shekhar2018anomalous}. One experimental signature of these Weyl nodes is the chiral anomaly: charge pumping between Weyl nodes of opposite chirality when an applied magnetic field \textbf{B} is parallel to the electric field \textbf{E} \cite{adler1969axial, liang2018experimental}. This appears as a negative longitudinal magnetoresistance (LMR) with a characteristic angle dependence $\mathbf{E} \cdot \mathbf{B}$, and has been observed in several $Ln$PtBi compounds \cite{chen2020chiral, hirschberger2016chiral, shekhar2018anomalous, liang2018experimental}.

A fundamental challenge, however, is that negative LMR is not unique to the chiral anomaly. Other mechanisms for negative LMR include current jetting \cite{liang2018experimental, dos2016search}, conductivity fluctuations \cite{hu2005current, dos2016search, steele1955anomalous, schumann2017negative}, and spin-disorder scattering in magnetically ordered materials \cite{usami1978magnetoresistance}. In half Heusler compounds, conductivity fluctuations may contribute because these materials are highly susceptible to natural nonstoichiometry \cite{yonggang2017natural} and variations in atomic site ordering \cite{ougut1995band}, due to the low formation energies for point defects. Moreover, since $Ln$PtBi and $Ln$PtSb are typically antiferromagnetic below $T_N \sim 10$ K, spin disorder scattering is likely to play a role.

Here we explore the magnetotransport properties of GdPtSb, which like GdPtBi, is seen in bandstructure calculations to have quadratic bands that touch at $Gamma$, taking into account the effects of naturally occurring nonstoichometry. Specifically, we examine effects of Pt vacancies on magnetotransport of GdPtSb films, grown by molecular beam epitaxy on c-plane sapphire substrates. We find that the angle-dependent magnetoresistance depends strongly on Pt stoichiometry, and comment on the relative roles of chiral anomaly, conductivity fluctuations, spin-disorder scattering, and topological hall effect in GdPt$_x$Sb. In particular, our observation of a plateau in the Hall resistivity in near stoichiometric GdPtSb suggests a topological Hall effect which may arise from topological spin textures that have not previously been observed in $Ln$PtSb or $Ln$PtBi systems.

\section{Results}

\begin{figure}[h]
    \centering
    \includegraphics[width=0.45\textwidth]{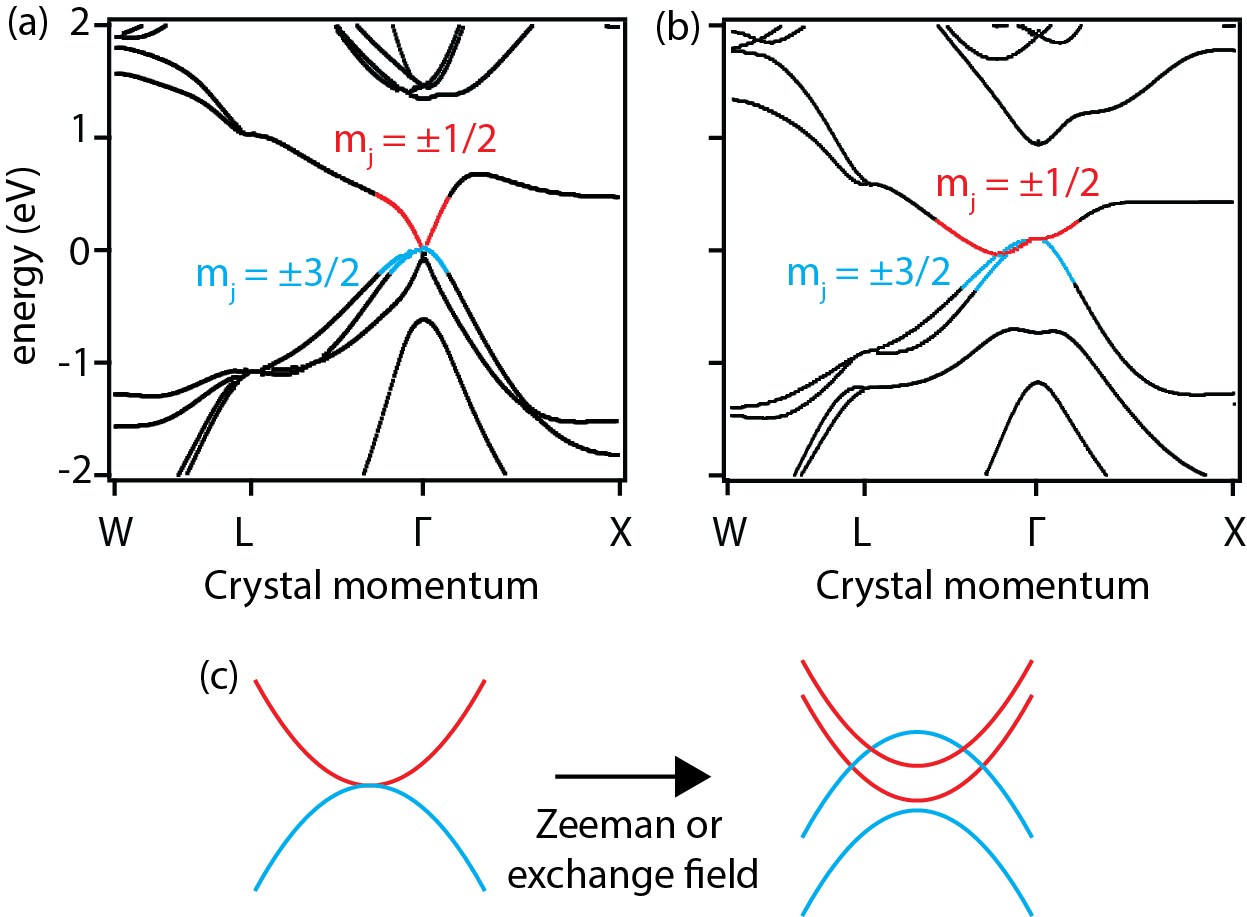}
    \caption{DFT calculations. (a) DFT-GGA band structures of GdPtSb and (b) GdPtBi. The bands near the Fermi energy at $\Gamma$ have strong Sb $5p_{3/2}$ or Bi $6p_{3/2}$ character. The red color denote the $m_j=\pm 1/2$ states and the blue color denote the $m_j=\pm 3/2$ states. (c) Cartoon showing how the quadratic band touching at $\Gamma$ splits due to exchange or Zeeman splitting to form pairs of Weyl nodes with opposite chirality.}
    \label{dft}
\end{figure}

We first establish the possibility of Weyl nodes for GdPtSb using density functional theory (DFT) calculations. In the well studied material GdPtBi, the essential feature is a quadratic band touching of four Bi $6p$ $j=3/2$ states near the Fermi energy (Fig. \ref{dft}(b)). The Zeeman energy or exchange energy splits the $m_j=\pm 3/2$ and $m_j=\pm 1/2$ to create Weyl nodes near the Fermi energy (Fig. \ref{dft}(c)) \cite{hirschberger2016chiral, shekhar2018anomalous} . 

Our nonmagnetic DFT calculations suggest that GdPtSb replicates this essential feature, with a quadratic touching of Sb $5p$ $j=3/2$ states at the Fermi energy at $\Gamma$ (Fig. \ref{dft}(a)). However, we note that for GdPtSb, there is an additional hole band approximately $60$ meV below the charge neutrality point (0 eV) at $\Gamma$ that may complicate the transport by providing an additional conduction channel. For GdPtBi, this hole band is pushed further down to $\sim 700$ meV, presumably due to the larger spin-orbit coupling for Bi compared with Sb.

\begin{figure}[h]
    \centering
    \includegraphics[width=0.45\textwidth]{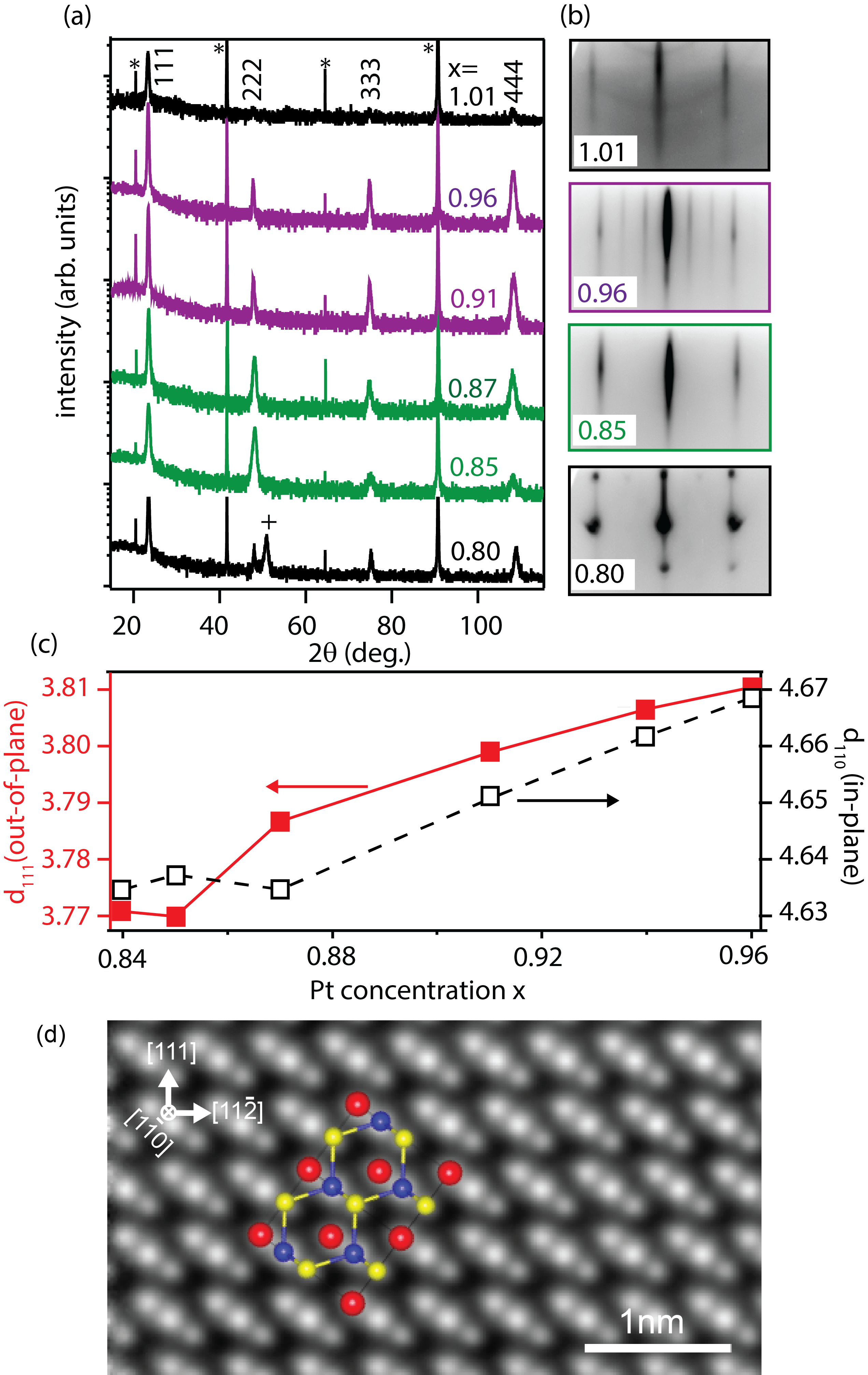}
    \caption{Structures and basic transport properties of stoichiometric and Pt-deficient GdPtSb epitaxial films. (a) X-ray diffraction (Cu $K\alpha$ of (111)-oriented GdPt$_x$Sb thin films grown on Al$_2$O$_3$ (0001). The ``+'' sign for $x=0.80$ denotes a GdSb impurity phase. (b) Corresponding reflection high energy electron diffraction (RHEED) patterns. (c) Out-of-plane ($d_{111}$, red) and in-plane ($d_{110}$, blue) lattice spacings extracted from on axis $222$ and off axis $202$ reflections (Supplemental). (d) High angle annular dark field scanning transmission electron microscopy (STEM) image of GdPtSb, measured along a $\langle 110 \rangle$ zone axis.}
    \label{structure}
\end{figure}

We synthesize GdPtSb films by molecular beam epitaxy on (0001)-oriented Al$_2$O$_3$ substrates, using conditions similar to Ref. \cite{du2019high}. The growth temperature is 600 $\degree$C. The Gd flux was supplied by a thermal effusion cell. A Sb$_2$/Sb$_1$ mixture was supplied by a thermal cracker cell with a cracker zone operated at 1200 $^\circ$C. The Pt flux was supplied by an electron beam evaporator. Fluxes were measured in-situ using a quartz crystal microbalance (QCM) immediately prior to growth. Absolute compositions were measured by Rutherford Backscattering Spectrometry (RBS). Due to the high relative volatility of Sb, GdPtSb films were grown in an Sb adsorption-controlled regime with a 30\% excess Sb flux, such that the Sb stoichiometry is self regulated \cite{shourov2020semi}.

X-ray diffraction (XRD) and reflection high energy electron diffraction (RHEED) measurements reveal that that epitaxial (111)-oriented GdPt$_x$Sb films with half Heusler structure are readily stabilized under Pt-deficient conditions (Fig. \ref{structure}(a)). For $x=0.85$ to 1, only the anticipated $111$-type reflections are observed by XRD indicating phase pure half Heusler growth. The corresponding streaky RHEED patterns indicate smooth epitaxial films. Samples closer to stoichiometry show an enhanced intensity of the higher order $333$ and $444$ XRD reflections and a well ordered $3\times$ surface reconstruction by RHEED, compared to the $1\times$ periodicity for Pt deficient films ($x=0.85$). For highly Pt deficient conditions, $x\leq 0.8$, we observe precipitation of a secondary GdSb phase by XRD and rough three-dimensional growth by RHEED. For Pt-rich conditions $x > 1.01$ the higher order $222$, $333$, and $444$ half Heusler XRD reflections disappear and we observe faint polycrystalline rings in the RHEED pattern. 

\begin{figure}[ht]
    \centering
    \includegraphics[width=0.45\textwidth]{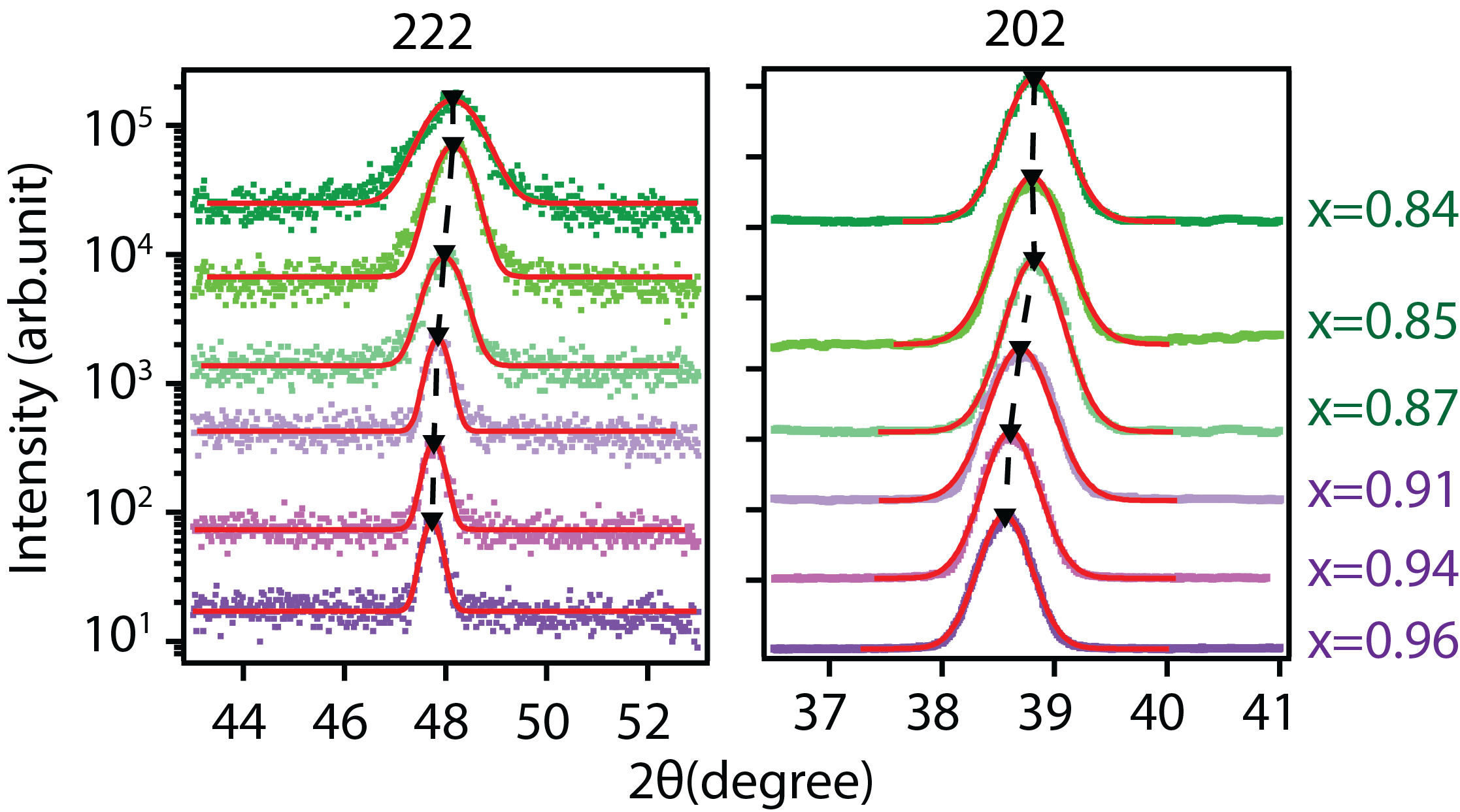}
    \caption{XRD $222$ and $202$ reflections used to extract the in plane and out of plane lattice spacings, respectively.}
    \label{supp_xrd}
\end{figure}

Focusing on the phase pure GdPt$_x$Sb samples with $x=0.85-1$, we observe a systematic increase for the out-of-plane $d_{111}$ and in-plane $d_{110}$ lattice spacings (Fig. \ref{structure}(c)). The in-plane $d_{110}$ spacings are calculated from measurements of an off-axis $202$ reflection (Appendix Fig. \ref{supp_xrd}). The observed Pt deficient samples are consistent with DFT calculations that predict Pt vacancies are the lowest energy defects for the related compound LuPtSb \cite{khalid2022defect}. High angle annual dark field (HAADF) scanning transmission electron microscopy (STEM) measurements of the $x=0.85$ sample shown in Fig. \ref{structure}(d) are in agreement with the expected site ordering for GdPtSb in cubic half Heusler structure (space group F$\bar{4}$3m). Here, the brightest atomic columns correspond to columns of Pt atoms (which have the largest atomic mass), followed by columns of Gd and columns of Sb.

\begin{figure}[h]
    \centering
    \includegraphics[width=0.47\textwidth]{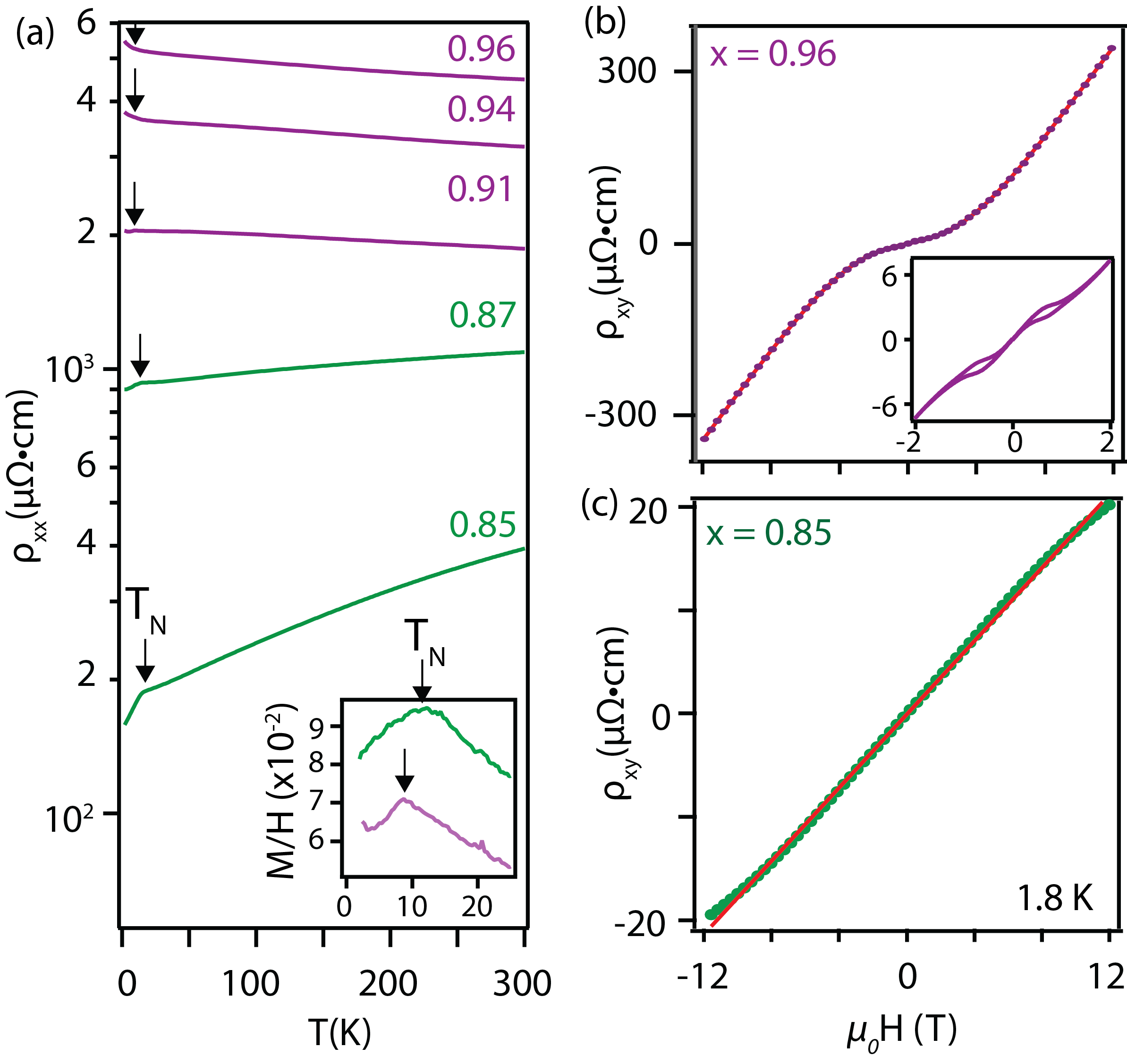}
    \caption{(a) Zero-field resistivity for GdPt$_x$Sb with different Pt concentration. Insert shows the magnetic susceptibility versus temperature for the $x=0.84$ (green) and $x=0.96$ (purple) samples measured by SQUID, showing the Neel transition. Kinks in the susceptibility coincide with kinks in the resistivity. (b) Transverse (Hall) resistivity for the $x=0.96$ sample, measured at 1.8 K. At low field $|B|<1$ T we observe nonlinearities suggestive of the topological Hall effect. Higher field ($|B|>2$ T) nonlinearities are associated with ordinary Hall effect with multiple carriers. The red line is an ordinary Hall effect two band model fit. (c) Hall effect at 1.8 K for the $x=0.85$ sample showing near linear behavior for $|B|<6$ T.}
    \label{transport}
\end{figure}

Zero-field resistivity measurements of near stoichiometric GdPtSb films ($x>0.9$) display an insulator-like temperature dependence ($d\rho/dT<0$, Fig. \ref{transport}(a) purple). We attribute the insulator-like resistivity to the Fermi energy being near the quadratic band touching (Fig. \ref{dft}). In contrast, for heavily Pt-deficient samples ($x<0.9$) we observe metallic transport ($d\rho/dT>0$, Fig. \ref{transport}(a), green). We attribute the more metallic transport to doping induced by Pt vacancies. 

\begin{figure}[h]
    \centering
    \includegraphics[width=0.47\textwidth]{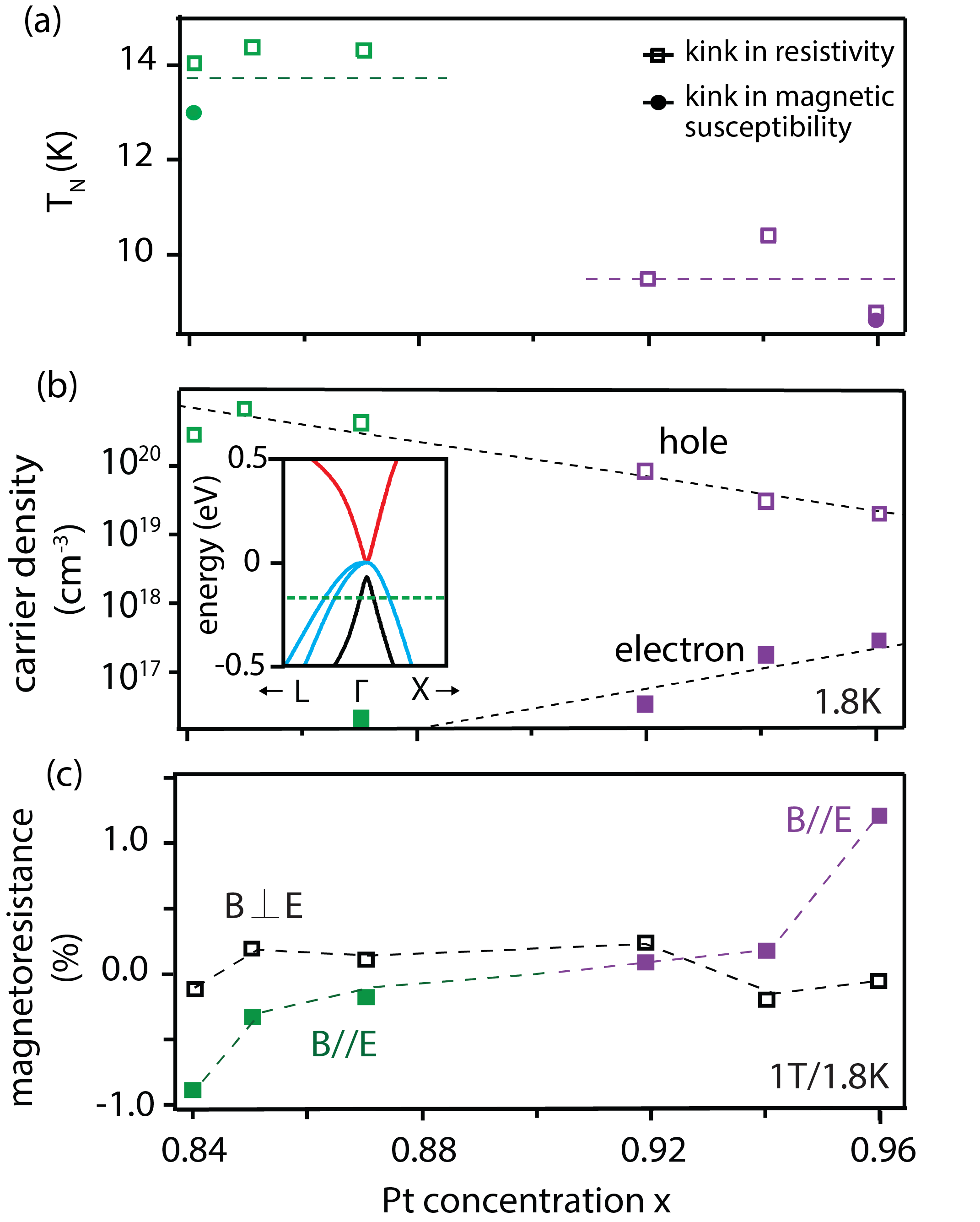}
    \caption{(a) Neel temperature ($T_N$) extracted from kinks in the temperature dependent magnetic susceptibility and resistivity. (b) Hall carrier density versus Pt concentration at 1.8 K, extracted from fitting to a model with one electron and one hole. The insert shows the approximate position of the Fermi energy for $x=0.85$ in a two band model, which is clearly an over simplification. (c) Longitudinal ($\mathbf{B} \parallel \mathbf{E}$) and transverse ($\mathbf{B} \perp \mathbf{E}$) magnetoresistance at 1 T and 1.8 K for samples with varying Pt concentration.}
    \label{summary}
\end{figure}

We observe kinks in the temperature dependent resistivity that correspond to kinks at the same temperature in magnetic susceptibility, as measured by superconducting quantum interference device (SQUID) magnetometry (Fig. \ref{transport}a insert). We attribute these kinks to the antiferromagnetic Neel transition $T_N$. We find that $T_N$ varies with Pt concentration $x$: the insulator-like samples ($x>0.9$) have a $T_N \approx 9$ K, whereas samples with a more metallic resistivity ($x<0.9$) have $T_N \approx 14$ K (Fig. \ref{summary}(a)). We speculate this jump in $T_N$ may arise from an enhanced Ruderman–Kittel–Kasuya–Yosida (RKKY) coupling between Gd moments and the Fermi sea with increasing carrier density.

Hall effect measurements reveal that Pt vacancies in GdPtSb are acceptors. The heavily Pt-deficient samples ($x<0.9$) show a positive and near linear dependence of the Hall resistivity $\rho_{xy}$ on magnetic field (Fig. \ref{transport}(c)), indicating dominant hole carriers. Closer to stoichiometry ($x>0.9$) the samples show stronger nonlinearities that are well fit by a two band model with one hole and one electron for $|B|>2$ T (Fig. \ref{transport}c). Fig. \ref{summary}(b) summarizes the effective electron and hole densities versus Pt concentration $x$, extracted from fitting to the two band model (Methods). Note that in reality there are 2-3 hole bands near the Fermi energy; however, since we are not able to distinguish these bands from a simple Hall effect fit we emphasize that these are ``effective'' carrier densities. We find that charge neutrality $n=p$ appears as $x$ approaches 1, and the effect of Pt vacancies is to increase the hole density and decrease the electron density. Based on a three-dimensional parabolic band model (Methods), we estimate that the Fermi energy for the $x=0.85$ sample, which has effective hole density $p = 2.72 \times 10^{20}$ cm$^{-3}$, lies 170 meV below the charge neutrality point. For the $x=0.96$ sample ($p=1.75 \times 10^{19}$ cm$^{-3}$, $n=2.5\times 10^{17}$ cm$^{-3}$), we estimate the Fermi energy is approximately 40 meV below the charge neutrality point. We caution that while we call this $x=0.96$ sample ``near stoichiometric,'' the hole density is still nearly two orders of magnitude larger than the electron density ($p >> n$).

Interestingly, the $x=0.96$ sample displays low field ($|B|< 1 $ T) hysteresis and nonlinearities consistent with the topological Hall effect (Fig. \ref{transport}(b), insert), suggesting a nontrivial Berry phase. This topological Hall effect often indicates topological spin textures like skyrmions, e.g., in chiral $B20$ compounds \cite{neubauer2009topological} and tetragonal Heusler compounds \cite{kumar2020detection}, but to our knowledge has not yet been reported in the $R$PtBi or $R$PtSb family. 

\begin{figure}[ht]
    \centering
    \includegraphics[width=0.45\textwidth]{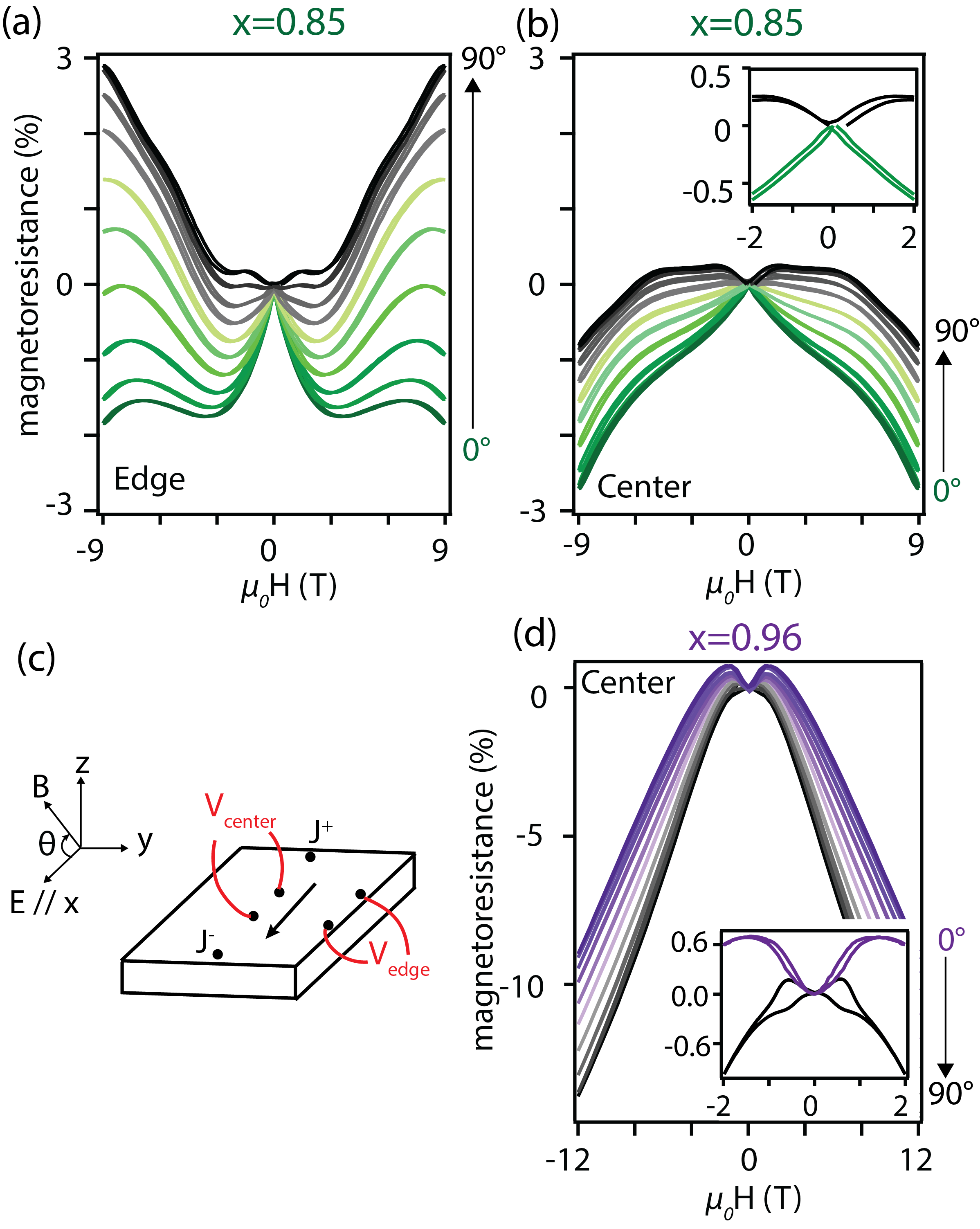}
    \caption{Chiral anomaly tests. (a,b) Magnetoresistance for a Pt deficient x=0.85 sample, as function of angle $\theta$ between $\mathbf{B}$ and $\mathbf{E}$. Comparison of voltage contacts along the edge (a) with voltage contacts along the center (b) constitutes the ``squeeze test'' for analyzing the contribution of extrinsic current jetting. (c) Measurement geometry. (d) Angle dependent magnetoresistance for a near stoichiometric $x=0.96$ sample. Insert shows hysteresis for $\theta=90 \degree$.}
    \label{chiral}
\end{figure}

We now analyze the magnetoresistance as a function of the angle $\theta$ between $\mathbf{B}$ and $\mathbf{E}$. The chiral anomaly, i.e. charge pumping between Weyl nodes, is expected to produce an additional current that is proportional to $\mathbf{B} \cdot \mathbf{E}$. Therefor, for Weyl semimetals the magnetoresistance $\Delta \rho / \rho_0$ should be negative for $\theta = 0$ and become more positive with increasing $\theta$ \cite{xiong2015evidence, parameswaran2014probing, hirschberger2016chiral, nielsen1983adler}. Fig. \ref{chiral}(a) shows the angle dependent magnetoresistance of the Pt deficient $x=0.85$ sample, measured at 1.8 K using voltage contacts along the edge of the sample and current sourced along the center of the sample (Fig. \ref{chiral}(c)). We find that the longitudinal magnetoresistance (LMR, $\theta = 0^\circ$) is negative and the transverse magnetoresistance (TMR, $\theta = 90 ^\circ$) is positive, as expected for the chiral anomaly. A continuously varying angular dependence, in a van der Pauw geometry, is shown in Appendix Fig. \ref{supp_theta}. The general angular dependence is qualitatively similar to previous studies of GdPtBi single crystals \cite{hirschberger2016chiral, liang2018experimental}. However, magnitude of change for our epitaxial GdPtSb films is only a few percent, whereas the magnetoresistance change for GdPtBi crystals is $\sim 80\%$.

\begin{figure}[ht]
    \centering
    \includegraphics[width=0.45\textwidth]{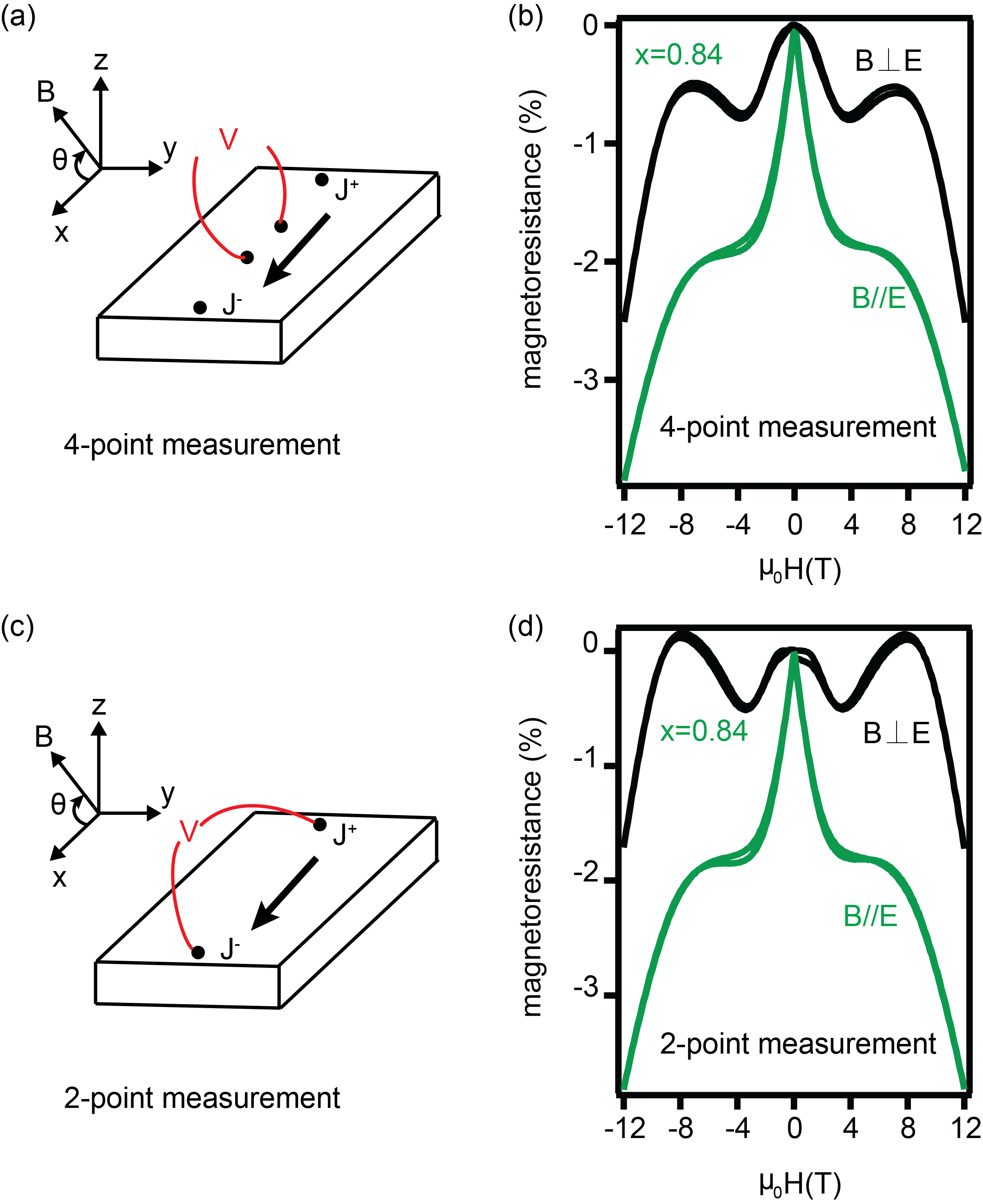}
    \caption{Two point versus four point resistivity test to analyze the contribution of current jetting.}
    \label{supp_2pt}
\end{figure}

The $x=0.85$ sample passes the ``squeeze test'' \cite{liang2018experimental}, suggesting that current jetting is not the primary origin of the negative LMR. Here, we find that measurements with contacts along the center of the sample (Fig. \ref{chiral}(b)) produce the same qualitative behavior as edge contacts (Fig. \ref{chiral}(a)), namely, negative LMR and positive TMR at modest magnetic field ($|B|< 6$ T). 2-point resistance measurements also produce the negative magnetoresistance for $\theta=0$ that increases with $\theta$ (Appendix Fig. \ref{supp_2pt}), confirming that current jetting is not a dominant factor. Furthermore, current jetting effects are expected to be strongest for materials with high carrier mobility and anisotropic conduction \cite{liang2018experimental}. Our samples have more modest Hall effect mobility ($\mu \sim 50$ cm$^2$/Vs) and are expected to be isotropic in the (111) plane, compared to reports of single crystals that have higher mobility of 1500 cm$^2$/Vs at 6 K \cite{hirschberger2016chiral}.

Near stoichiometric samples display the opposite angular dependence. Fig. \ref{chiral}(d) shows the angle-dependent magnetoresistance for a sample with composition $x=0.96$, measured at 1.8 K. Besides a weak positive magnetoresistance at low field ($B < 4$ T) and low $\theta$, the magnetoresistance is generally negative and becomes more negative with increasing $\theta$. This is opposite the expected dependence for chiral anomaly, which is expected to produce the most negative magnetoresistance for $\theta = 0$. Additionally, the transverse magnetoresistance shows a weak hysteresis within the range $|B| < 1$ T (Fig. \ref{chiral}(d) insert), the same range as observed for the Hall resistance (Fig. \ref{transport}(b) insert). Minimal hysteresis is observed in the magnetoresistance of the $x=0.85$ sample.

We summarize the TMR and LMR at field 1 T and temperature 1.8 K for several samples with varying Pt concentration in Fig. \ref{summary}(c). For Pt deficient samples ($x<0.9$) the LMR ($\mathbf{B} \parallel \mathbf{E}$) is generally more negative than the TMR ($\mathbf{B} \perp \mathbf{E}$), as expected for chiral anomaly. For samples closer to stoichiometry ($x>0.9$) the LMR is generally more positive than TMR. This dependence on Pt concentration is opposite to the expectation for chiral anomaly, which we expect to be strongest for stoichiometric samples in which the Fermi energy is closer to the Weyl nodes ($x \approx 1$). Previous experiments on doped GdPtBi show that the negative LMR is maximized near the charge neutrality point \cite{hirschberger2016chiral}. 

\begin{figure}[ht]
    \centering
    \includegraphics[width=0.45\textwidth]{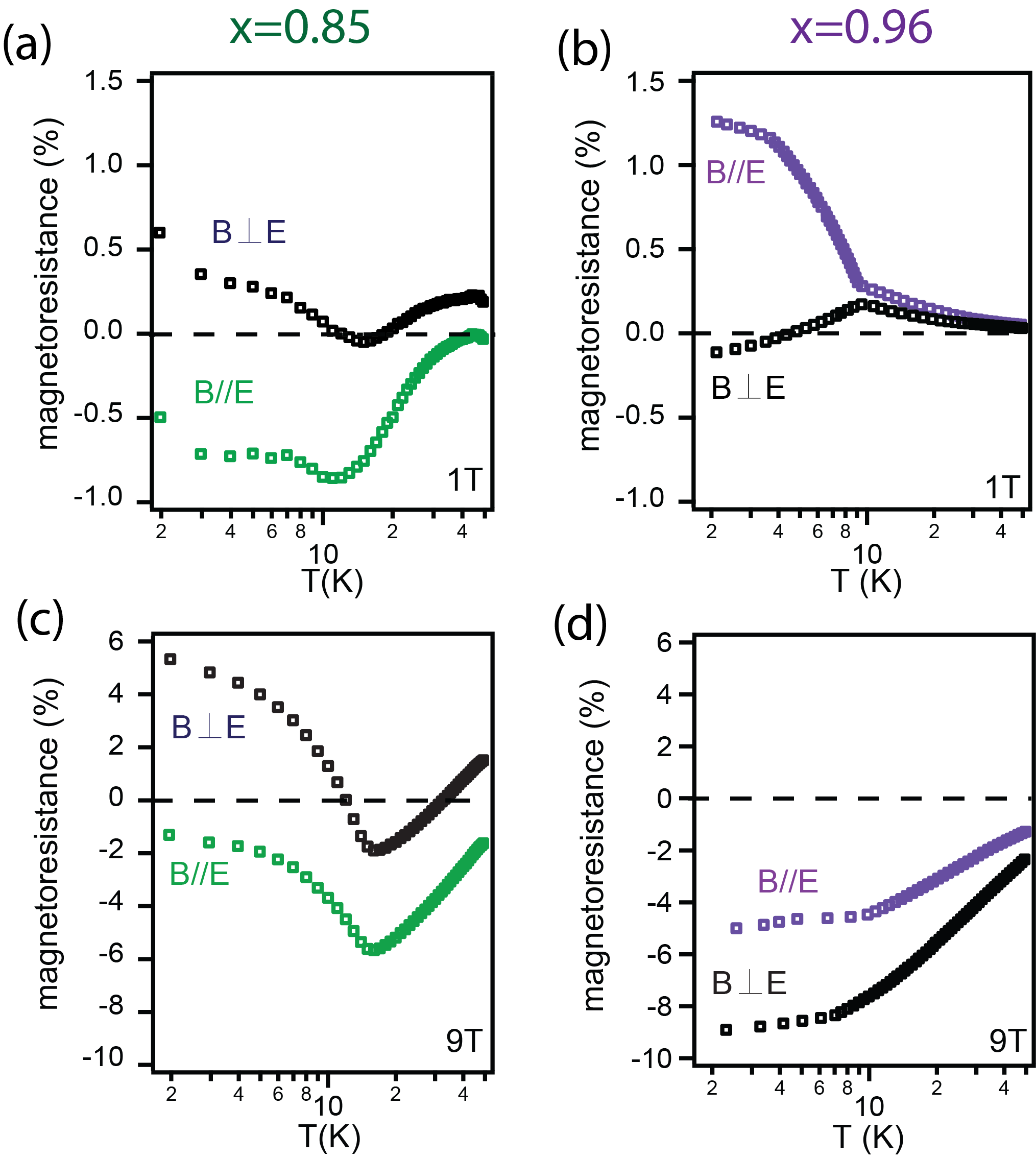}
    \caption{Temperature dependence of the LMR and TMR. (a) $x=0.85$ sample at 1 T. (b) $x=0.95$ sample at 1 T. (c) $x=0.85$ sample at 1 9 T. (e) $x=0.95$ sample at 9 T. These measurements were performed in a van der Pauw geometry. Note the magnitudes are different than in Fig \ref{chiral} due to the different contact geometry.}
    \label{MR_T}
\end{figure}

\section{Discussion}

What explains the Pt stoichiometry dependence? For Pt deficient samples ($x<0.9$), the Pt vacancy disorder and position of the Fermi energy away from the Weyl nodes suggests that conductivity fluctuations rather than chiral anomaly may dominate. For inhomogeneous or disordered materials, spatial fluctuations in the conductivity can lead to a component of the carrier velocity that is perpendicular to $\mathbf{B}$, even when the global $\mathbf{E}$ is parallel to $\mathbf{B}$ \cite{hu2005current, hu2007nonsaturating, schumann2017negative, dos2016search, xu2019negative}. The resulting Lorentz force leads to a decrease in the LMR with increasing $\mathbf{B}$. We expect Pt-deficient samples to show these fluctuations more strongly than stoichiometric samples. Moreover, the very Pt deficient samples have Fermi energy furthest away from the expected Weyl nodes, and thus are not anticipated to show strong effects from chiral anomaly compared to samples closer to stoichiometry ($x \approx 1$).

For samples closer to stoichiometry, trivial bands near the Fermi energy and spin disorder scattering may explain the negative LMR. We first note that unlike the well studied Weyl semimetal GdPtBi, GdPtSb has an additional hole band approximately 60 meV below the charge neutrality point (Fig. \ref{dft}(a)) that may contribute to the transport and obscure the chiral anomaly. 

Additionally, in magnetically ordered materials like GdPtSb (which is antiferromagnetic), field alignment of spins or other forms of spin-disorder scattering can also cause a negative LMR \cite{usami1978magnetoresistance, ritchie2003magnetic,borca2001evidence}, and the angular dependence can arise from magnetocrystalline anisotropy or shape anisotropy. For reasons that are still unclear, the effects of magnetic ordering are more prominent for our near stoichiometric GdPtSb samples than for heavily Pt-deficient samples. First, the TMR of the $x=0.96$ sample shows magnetic hysteresis that does not appear for the $x=0.85$ sample (Fig. \ref{chiral}(b,d) inserts). Second, the temperature-dependent magnetoresistance for the $x=0.96$ sample at low field diverges below $T_N \sim 9$ K (Fig. \ref{MR_T}(b)), where the LMR increases and the TMR decreases with decreasing temperature. In contrast, the LMR and TMR for the $x=0.85$ sample do not diverge below $T_N$ (Fig. \ref{MR_T}(a)). Third, $T_N$ abruptly jumps from $\sim 14$ K for Pt deficient samples to $T_N \sim 9$ K for near stoichiometric samples (Fig. \ref{summary}(a)). This suggests an abrupt change in the exchange, and possibly magnetic ordering, as a function of $x$. Further magnetization studies, as a function of magnetic field orientation, are required to understand the mechanisms for the angle-dependent magnetoresistance in near stoichiometric samples.

\begin{figure}[ht]
    \centering
    \includegraphics[width=0.45\textwidth]{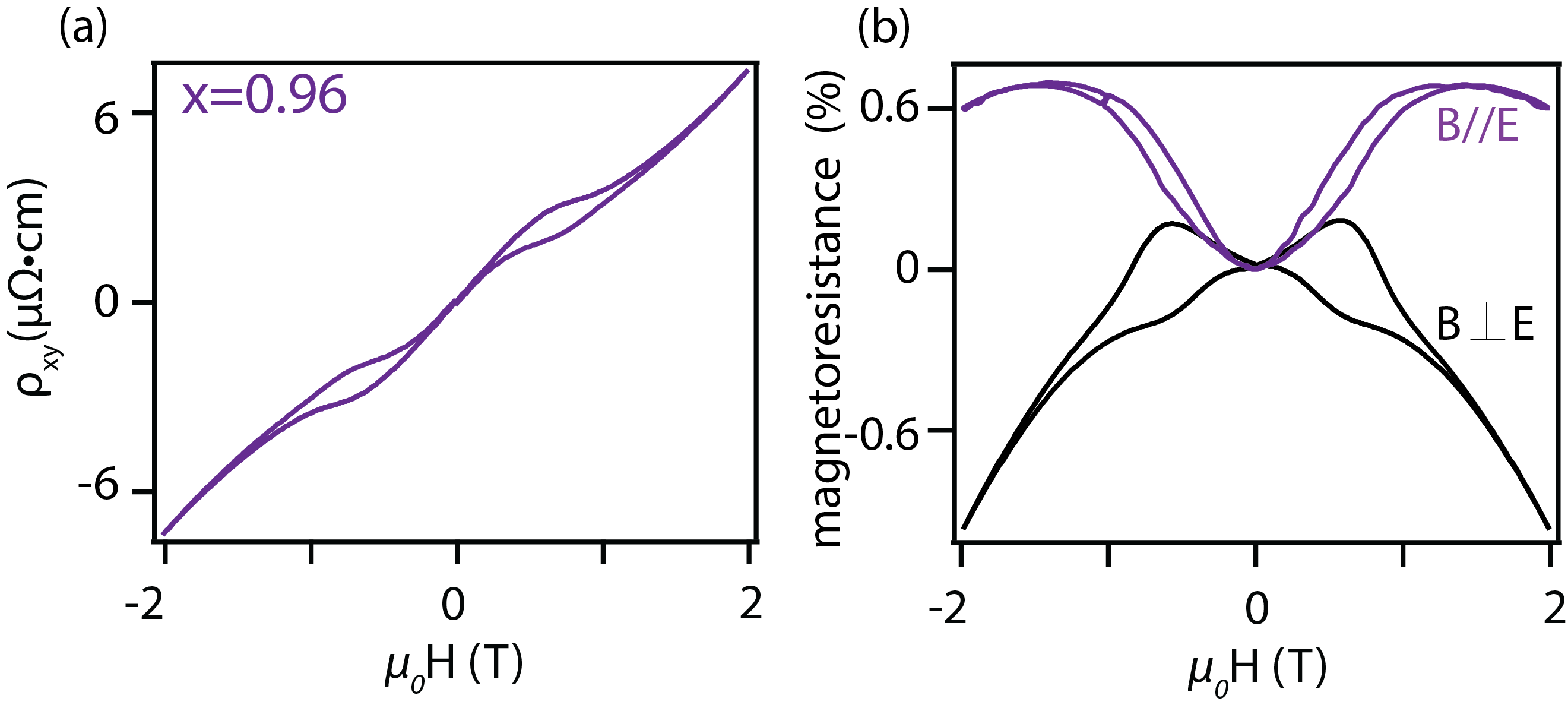}
    \caption{(a) Low field Hall effect, showing possible topological hall effect. No background has been subtracted. (b) Low field magnetoresistance showing hysteresis over the same range as the possible topological component of the Hall effect ($|B|<1$ T).}
    \label{topological}
\end{figure}

More intriguingly, the near stoichiometric samples display indications of topological Hall effect (Fig. \ref{transport}b insert) that do not appear for the very Pt deficient samples (Fig. \ref{transport}(a,b)). We reproduce the low field transverse (Hall) and longitudinal magnetroresistance of the $x=0.96$ sample in Fig. \ref{topological}. In the transverse resistivity $\rho_{xy}$ versus magnetic field, we observe low field nonlinearities and hysteresis that is reminiscent of the topological Hall effect. Topological Hall effect indicates a nontrivial Berry phase and is often an indication of chiral spin textures like skyrmions; however, a more detailed analysis is required to clearly identify this nonlinearity as topological Hall effect \cite{gerber2018interpretation}. The low field magnetoresistance (Fig. \ref{chiral}(d) insert and Fig. \ref{topological}(b)) shows the same width of hysteresis ($\pm 1$ T). Systems that display topological Hall effect also display negative magnetoresistance \cite{ahadi2017evidence, huang2022plateau}.

\section{Conclusions}

We demonstrate that half Heusler GdPt$_x$Sb epitaxial films grown on sapphire accommodate a large concentration of Pt vacancies, which act as acceptors. Samples with high Pt vacancy concentration have a metallic resistivity vs temperature and display a negative longitudinal magnetoresistance that becomes more positive as the magnetic field tilts away from the electric field, consistent with chiral anomaly. However, the large concentration of Pt vacancies and position of the Fermi energy away from the Weyl nodes suggests that conductivity fluctuations, rather than chiral anomaly, dominates. Samples closer to stoichiometry show an insulator-like resistivity versus temperature and TMR that is more negative than LMR, opposite the expected behavior of chiral anomaly. The low field Hall effect shows nonlinearities similar to topological Hall effect. Further detailed magnetotransport studies are required to understand the possible balance of Weyl nodes and topological spin textures in GdPtSb.

\section{Methods}

\textbf{First-principles calculations.} Calculations were done with ABINIT using the PBE GGA exchange correlation potential and norm-conserving pseudopotentials from  ONCVPSP-3.3.0, except for the Gd pseudopotential, which was constructed to have the f-orbitals in the core. The k point mesh used was $18 \times 18 \times 18$ and the energy cutoff was 1400 eV (50 Hartree). Computed lattice parameters for the half Heusler structure were 6.647 \AA\ for GdPtSb and 6.777 \AA\ for GdPtBi in the conventional fcc unit cell.

\textbf{Transport measurements and fitting.} Magnetotransport measurements for GdPt$_x$Sb samples were performed using a Quantum Design Dynacool Physical Property Measurement System. Hall effect measurements were generally performed in a Hall bar geometry with typical dimensions 5 mm by 1 mm. For the $x=0.85$ sample the Hall effect was measured in a van der Pauw geometry. Angle-dependent magnetoresistance measurements were performed using a horizontal rotator probe and contacts in a ``squeeze test'' geometry, as shown in Fig. \ref{chiral}(c). The ``center'' channel is a standard linear four point geometry with contacts in a line down the center of the sample, where the outer contacts are current and the inner contacts are voltage. The ``edge'' channel uses the same current contacts, but places the voltage contacts at the edge of the sample to test the effects of current jetting down the center of the sample.

Nonlinear Hall data are fit with a two-band model of the following form
\begin{equation*}
\rho_{xy}(B)=\frac{B}{e}\frac{(n_h\mu_h^2-n_e\mu_e^2)+(n_h-n_e)\mu_h^2\mu_e^2B^2}{(n_h\mu_h+n_e\mu_e)^2+(n_h-n_e)^2\mu_h^2\mu_e^2B^2}
\end{equation*}
where $n_h$ ($n_e$) and $\mu_h$ ($\mu_e$) are concentration and mobility of holes (electrons). We constrain the two-band Hall fit by also fitting to the zero magnetic field longitudinal resistivity
\begin{equation*}
\rho_{xx}(0)=\frac{1}{e(n_h\mu_h+n_e\mu_e)}.
\end{equation*}

We use the carrier concentration to estimate the Fermi level positions. For a rough estimation, we used 3D parabolic band density of states.
\begin{equation*}
D(E)=\frac{\sqrt2}{\pi^2}\frac{(m^*)^{\frac{3}{2}}}{\hbar^3}\sqrt{E}
\end{equation*}
where m$^*$ is the effective mass.
\begin{equation*}
m^*=\frac{\hbar^2}{d^2E/dk^2}
\end{equation*}
We calculated the effective mass for the 3 valence bands near E=0 eV at $\Gamma$ point, based on our band structure calculation in Fig. \ref{dft}.
The Fermi level is extracted from the following integral,
\begin{equation*}
\sum_{n=1}^{2 or 3}\int_{0}^{E_F} D(E) \,dE=\frac{N}{V} 
\end{equation*}
From the above calculation, the E$_F$ for the $x=0.84$ sample with hole concentration $2.72 \times 10^{20}$ cm$^{-3}$ is about -170 meV and that for the $x=0.96$ sample with hole concentration 
 $1.75 \times 10^{19}$ cm$^{-3}$ is about -40 meV.

\textbf{Magnetization measurements.} 
Magnetic properties were measured using a Quantum Design MPMS SQUID (Superconducting Quantum Interference Device) Magnetometer. The magnetic field is applied perpendicular to the sample surface. The net magnetization data for the GdPt$_x$Sb thin films is extracted by subtracting a background measurement of the Al$_2$O$_3$ substrate from the total magnetization signal.

\textbf{Transmission electron microscopy.} 
GdPtSb cross-section samples were prepared with a Zeiss Ga-focused ion beam, followed by final thinning in a Fishione Model 1040 Nanomill using Ar ions at 900V. Samples were stored in vacuum and cleaned in a GV10x DS Asher cleaner at 20 W for 10 min to remove contamination and minimize the oxidization on the sample before being transferred into the STEM column. A probe corrected Thermo-Fisher Titan STEM equipped with CEOS aberration corrector operated at 200 kV was used to collect the atomic resolution STEM images. A 24.5 mrad probe semi-angle, 18.9 pA probe current was used to collect HAADF image series with a Fishione 3000 annular detector covering collection angle ranging from 53.9 to 269.5 mrad. Each frame in the image series took about 0.6 second to acquire with 10 $\mu$s on each STEM scan position and 256-by-256 scan grid. Non-rigid registration \cite{yankovich2014picometre} was used to compensate for the drift and distortions during image series acquisition, before the series was averaged to get one single frame with high signal-to-noise ratio as shown in Fig. \ref{structure}(e).

\textbf{X-ray diffraction.} X-ray diffraction measurements were performed using a Malvern Empyrean diffractometer with Cu $K\alpha$ radiation.

\textbf{Rutherford Backscattering Spectrometry (RBS).} RBS measurements were performed at the University of Minnesota Characterization Facility.

\section{Acknowledgment}

We thank Max Hirschberger for helpful discussions on chiral anomaly tests. We thank Greg Haugstad for performing RBS measurements. Special thanks to D. R. Hamman for providing the Gd pseudopotential.

Heusler epitaxial film growth and magnetotransport at the University of Wisconsin were supported by the Air Force Office of Scientific Research (FA9550-21-0127). Preliminary synthesis was supported by the Army Research Office (ARO Award number W911NF-17-1-0254) and the National Science Foundation (DMR-1752797). Calculations by KR and KG were supported by the Office of Naval Research N00014-21-1-2107. Transport measurements at the University of Minnesota by L.R.T. and B.J. were supporteb by the National Science Foundation through the University of Minnesota MRSEC under Award No. DMR-2011401. TEM experiments by CZ and PMV were supported by the US Department of Energy, Basic Energy Sciences (DE-FG02-08ER46547) and used facilities are supported by the Wisconsin MRSEC (DMR-1720415). We gratefully acknowledge the use of x-ray diffraction facilities supported by the NSF through the University of Wisconsin Materials Research Science and Engineering Center under Grant No. DMR-1720415. We acknowledge PARADIM for annealed sapphire substrates. 

\bibliographystyle{apsrev}
\bibliography{ref}

\clearpage \newpage

\section{Appendix}

\begin{figure}[ht]
    \centering
    \includegraphics[width=0.45\textwidth]{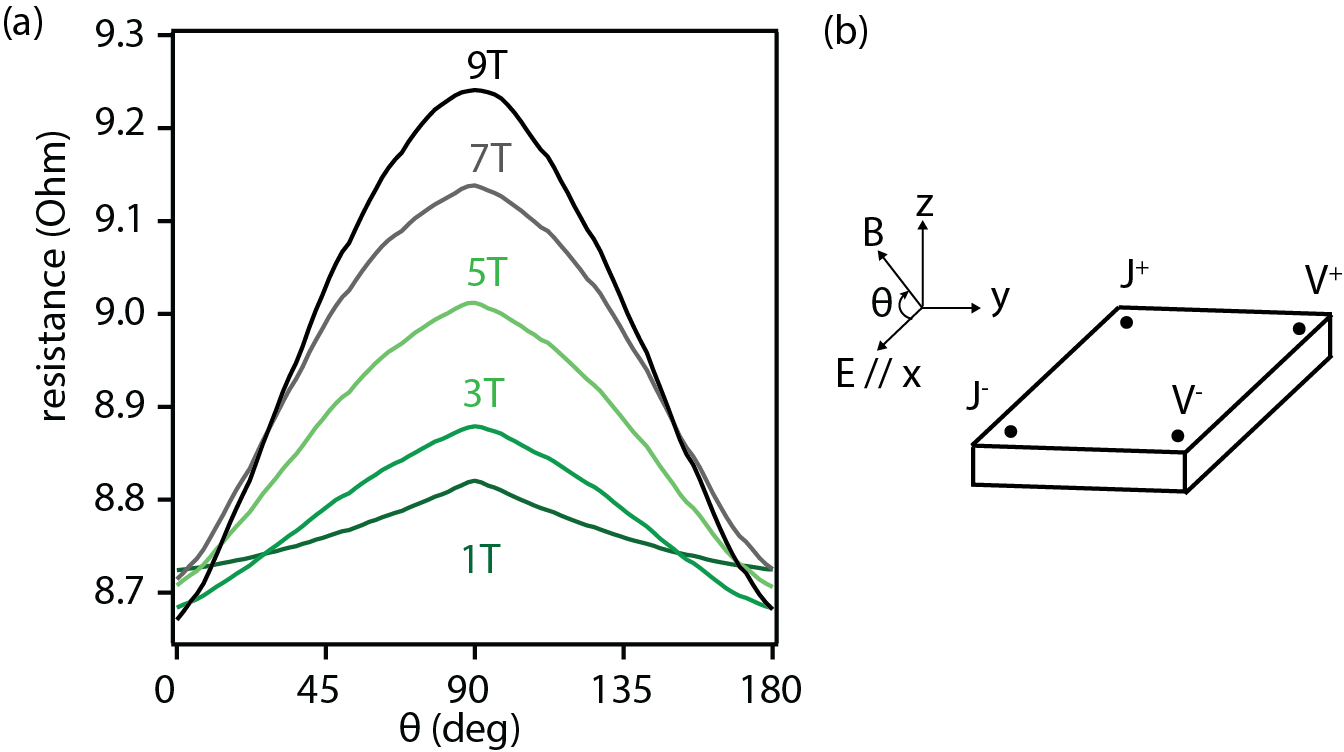}
    \caption{(a) Magnetoresistance versus $\theta$ for the $x=0.85$ sample. (b) This measurement was performed in a van der Pauw geometry.}
    \label{supp_theta}
\end{figure}

\end{document}